\documentclass[Universe,article,submit,pdfLaTex,oneauthor]{Definitions/mdpi} 

\firstpage{1} 
\pubvolume{1}
\issuenum{1}
\articlenumber{0}
\pubyear{2024}
\copyrightyear{2024}
\datereceived{Aug. 12 2024} 
\dateaccepted{Sept. 2, 2024} 
\datepublished{Sept. 5. 2024} 
\hreflink{https://doi.org/} 
\Title{Non-Canonical Dark Energy Parameter Evolution in a Canonical Quintessence Cosmology}

\TitleCitation{Special Issue for the Journal Universe "[Dark Energy and Dark Matter]" }

\Author{Rodger I. Thompson $^{1,\dagger,\ddagger}$\orcidA{}*}

\AuthorNames{Rodger I. Thompson}
\AuthorCitation{Thompson, R. I.}

\address{$^{1}$ \quad Department of Astronomy and Steward Observatory University of Arizona; Tucson, AZ. 85721,USA; 
rit@arizona.edu}

\corres{Correspondence: rit@arizona.edu;}
\nolinenumbers

\abstract{This study considers the specific case of a flat, minimally coupled to gravity, quintessence cosmology with a dark energy 
quartic polynomial potential that has the same mathematical form as the Higgs potential.  Previous work on this case determined
that the scalar field is given by a simple expression of the Lambert W function in terms of the easily observable scale factor.  This
expression provides analytic equations for the evolution of cosmological dark energy parameters as a function of the scale factor
for all points on the Lambert W function principal branch.  The Lambert W function is zero at a scale factor of zero that marks the
big bang.  The evolutionary equations beyond the big bang describe a canonical universe that is similar to $\Lambda$CDM making
it an excellent dynamical template to compare with observational data.  The portion of the W function principal before the big
bang extends to the infinite pre bang past.  It describes a noncanonical universe with an initially very low mass density that is
contracting by rolling down the dark energy potential to a singularity, big bang, at the scale factor zero point.  This provides a
natural origin of the big bang.  It also raises the possibility that the universe existed before the big bang and is far older and was
once far larger than its current size.  The recent increasing interest in the possibility of a dynamical universe instead of
$\Lambda$CDM makes the exploration of the nature of such universes particularly relevant.} 


\begin{document}

\section{Introduction}\label{s-intro} 
Flat, minimally coupled to gravity quintessence was one of the first and is probably one of the simplest rolling scalar field cosmologies.
Even so quintessence has a complex range of properties that depend on the mathematical form of the dark energy potential and its
arguments, as well as the boundary conditions on the cosmological parameters such as the current dark energy equation of state 
$w_0$.  The dark energy potential in this study is a quartic polynomial with the same mathematical form as the Higgs potential that 
is hence referred to as the Higgs Inspired, HI potential.  In a Ratra Peebles format  \cite{rat88, pee88} the HI potential is 
\begin{equation}\label{eq-HIpot}
V(\kappa\theta) = M^4((\kappa\theta)^2-(\kappa\delta)^2)^2 = M^4((\kappa\theta)^4-2(\kappa\theta)^2(\kappa\delta)^2+(\kappa\delta)^4)
\end{equation}
where $M$ is a constant with units of mass in terms of the reduced Planck mass $M_p$, and  $\kappa$ is the inverse reduced mass
$\frac{1}{M_p}$ The scalar field is $\theta$ and $\delta$ is a constant, both with units of mass, which makes $\kappa\theta$ 
and $\kappa\delta$ dimensionless.  All of the HI potential dimensionality is in the $M^4$ term. Previous work \cite{thm23} on this 
combination of cosmology and potential determined that the scalar field $\theta(a)$ has a simple solution in terms of the scale factor 
$a$ and the Lambert W function \cite{olv10}.
\begin{equation}\label{eq-wa}
\kappa\theta(a) =\kappa\delta\sqrt{-W(\chi(a))}.
\end{equation}
The argument $\chi(a)$ of the W function is a power law function of the scale factor $a$.
\begin{equation}\label{eq-chi}
\chi(a)=q a^p.
\end{equation}
Both $q$ and $p$ are constants determined by the boundary conditions and the HI dark energy potential. The expression 
$(\kappa\theta)^2-(\kappa\delta)^2$ appears in the HI potential and several other places in the evolutionary 
functions. In terms of the W function it is.
\begin{equation}\label{eq-tw}
(\kappa\theta)^2-(\kappa\delta)^2 =- (\kappa\delta)^2(W(\chi(a))+1).
\end{equation}
Using Eqn.~\ref{eq-tw} the HI potential in the W function format is
\begin{equation}\label{eq-HIWpot}
V(a)= (M\kappa\delta)^4(W(\chi(a))+1)^2=(M\kappa\delta)^4(W(\chi(a))^2+2W(\chi(a))+1)
\end{equation}
The combination of the Lambert W function scalar, HI potential and the Quintessence cosmology is hereinafter referred to as the 
LHQ cosmology and universe.

\begin{figure}
    \centering
    \fbox{\includegraphics[width=8cm]{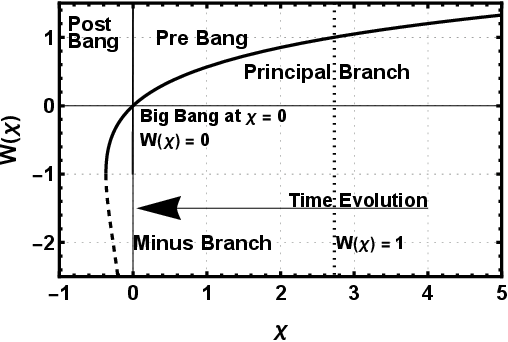}}
    \hspace{0.5cm}
    \fbox{\includegraphics[width=8cm]{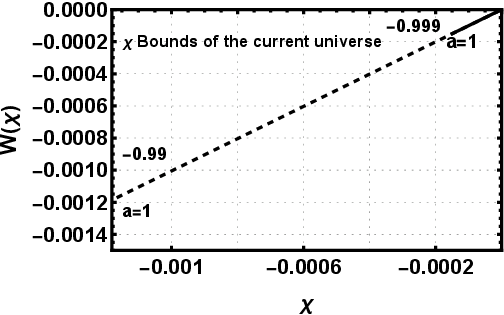}}
    \caption{\label{fig-W} The upper panel shows portions of the principal, solid line, and minus, dashed line, branches of the 
Lambert $W$ function $W(\chi)$. The big bang lies at the 0,0 origin. Evolution is from right to left as indicated by the arrow in 
the figure.  The dotted vertical line in the pre bang region marks the pre bang $W(\chi(a))=1$ point.  The slightly thicker portion 
of the $\chi=0$ axis indicates the width of the post bang universe. The lower panel shows the post bang universe.  The extent
of the solid line shows the present universe for $w_0=-0.999$ and the extent of the dashed line for $w_0=-0.99$.  The terminator
of each line is at a scale factor of one.}
\end{figure}

Figure ~\ref{fig-W} shows 
small portions of the W function principal and minus branch on both sides of the big bang.  
Although the Lambert W function minus branch is not considered in this work it is included to show that the maximum negative value 
of $\chi(a)$ lies at the end of the principal branch and beginning of the minus branch.  At the terminal point of the principal branch
$\chi=-\frac{1}{e}$ and $W(\chi)=-1$.

Figure~\ref{fig-W} displays many of the unique aspects of the LHQ cosmology and its evolution.  In keeping with its definition
as the ratio of the size of the universe relative to the current size set as one, the scale factor $a$ is positive on either side of
the big bang.  As shown later in section~\ref{sss-desc} $q$ is negative in the post bang region and positive in the pre bang region,
making $\chi$ negative in the post bang and positive in the pre bang regions. Figure~\ref{fig-W} shows that The Lambert W function
has the same sign as $\chi$.  This, in turn, indicates that the scalar field is real and positive in the post bang region and imaginary
in the pre bang region.  All of the pre bang cosmological parameters are, however, real.  This divides the LHQ cosmology universe into 
a canonical post bang era containing the canonical universe and the noncanonical pre bang era.    Although only plotted out to a 
pre bang $\chi$ of 5 the W function principal branch extends to infinity and the equation for the scalar is mathematically valid at all
points on principal branch.

A recent publication by the DESI collaboration \cite{des24} has claimed evidence for a preference of a dynamical cosmology over 
$\Lambda$CDM, drawing more interest to dynamical universe models.  Several publications, both positive and negative, have 
appeared relative to the DESI results, eg.\cite{muk24, roy24, jia24, orc24, pou24, gia24, din24}, employing parametric analysis, 
except for \cite{din24}.  This highlights the need for accurate, analytic, solutions based on physics, rather than parameterizations, 
for the evolution of dark energy cosmological parameters. 

The investigation centers on two regions of the principal branch.  The first is the far past in the pre bang epoch that extends to 
the infinite past. The study adopts a more practical starting point at a pre bang scale factor of 1000.  The region between a pre 
bang scale factor of one and 1000 is the far past region.  The region between a post bang scale factor of one and a pre bang scale 
factor of one, straddling the big bang is the second region, designated the transition zone.  In comparison to the far past the 
transition zone is miniscule, however, it includes all of the known post bang universe, the big bang, and the pre bang 
approach to the big bang.  

The existence of analytic simple equations for the dark energy scalar and potential provides the primary motivation for this 
investigation. These equations along with the quintessence cosmology, reviewed in Sect.~\ref{s-mcq} and the Friedmann
constraints provide the tools to calculate mathematically deterministic solutions for cosmological parameters as functions of the 
scale factor at all points on the W function principal branch.  The evolution of the dark energy parameters is determined by the 
principal branch track of the Lambert W function illustrated in Fig.~\ref{fig-W}. The matter and radiation average densities are 
strictly functions of their boundary condition densities and the scale factor.  

The remainder of the manuscript is arranged as follows.  After this introduction the basic quintessence equations are  reviewed in 
Sect.~\ref{s-mcq}. The mathematical properties of the cosmology are then established in terms of the Lambert W function in 
Sect.~\ref{s-mpc}.  These properties are the equations for the evolution of cosmological parameters as analytic functions of the scale 
factor.  The boundary conditions are set in Sect.~\ref{s-bc}.  The  physical properties for the far past and transition zone are 
established in Sec.\ref{s-ppc}.  These contain both mathematical and graphical representations of the evolution of 
cosmological parameters.  A set of conclusions are presented in Sect.~\ref{s-con}. The study uses natural units with 
$c$, $\hbar$ and $8 \pi G$ set to one. The unit of mass is the reduced Planck mass $M_p$. In these units $\kappa=1$ but is 
retained in equations to show the proper units. 

\section{Flat Minimally Coupled Quintessence}\label{s-mcq}
Even though they are well known, some of the more useful quintessence equations are presented here for easy reference.  The 
equations are expressed in the familiar form with $\phi$ as the true scalar.  The true scalar is $\phi = M \kappa \theta$ for the 
Ratra Peebles scalar $\kappa\theta$ used in this work.  Equations~\ref{eq-wa} and \ref{eq-HIpot} plus the equations in this
section, the Friedmann constraints in Sec.~\ref{ss-fc}, along with the boundary conditions in Sec~\ref{s-bc} form a complete
set of tools to produce deterministic solutions for the evolution of the primary cosmological parameters.

The action is
\begin{equation} \label{eq-act}
S=\int d^4x \sqrt{-g}[\frac{R}{2}-\frac{1}{2}g^{\mu\nu}\partial_{\mu}\partial_{\nu}\phi -V(\phi)] +S_m+S_r
\end{equation}
where $R$ is the Ricci scalar, $g$ is the determinant of the metric $g^{\mu\nu}$ and $V(\phi)$ is the dark energy potential. 
$S_m$ and $S_r$ are the actions of the matter and radiation fluids.  By definition the big bang has not occurred in the pre bang 
epoch so the radiation term is zero in the pre bang action.  The dark energy equation of state is $\frac{P_{\phi}}{\rho_{\phi}}$ 
with a dark energy density and pressure of
\begin{equation} \label{eq-rhop}
\rho_{\phi} \equiv \frac{\dot{\phi}^2}{2}+V(\phi), \hspace{1cm} P_{\phi}  \equiv \frac{\dot{\phi}^2}{2}-V(\phi).
\end{equation}
The kinetic term $\frac{\dot{\phi}^2}{2}$ is often referred to as $X$.
It follows that
\begin{equation}\label{eq-pprho}
P_{\phi} + \rho_{\phi}= \dot{\phi}^2, \hspace{0.5cm}\frac{P_{\phi} + \rho_{\phi}}{\rho_{\phi}}= (w+1)
\end{equation}
which provides the useful relation
\begin{equation}
\dot{\phi}^2 = \rho_{\phi}(w+1).
\end{equation}
The average matter and radiation densities are simply
\begin{equation}\label{eq-sm}
\rho_m=\frac{\rho_{m_0}}{a^3}, \hspace{0.5cm} \rho_r=\frac{\rho_{r_0}}{a^4}
\end{equation}
where $\rho_{m_0}$ and $\rho_{r_0}$ are the matter and radiation densities at a scale factor of one. Only these average 
densities are considered in this study. 

\section{The Mathematical Properties of The LHQ Universe}\label{s-mpc}
This section develops the analytic functions for the evolution of cosmological parameters in both the pre and post bang epochs.   
Some duplication of derivations in \cite{thm23} occurs but are included in order to have most derivations in this manuscript rather 
than having to refer back to the previous publication.

\subsection{Friedmann constraints}\label{ss-fc}
The first and second Friedmann constraints, given below, are important tools in deriving the cosmological parameter evolution
equations.
\begin{equation}\label{eq-f1}
3H^2= \rho_{de} +\rho_m +\rho_r.
\end{equation}
\begin{equation}\label{eq-f2}
3(\dot{H} + H^2) = -\frac{\rho_{de} +\rho_m +\rho_r +3P}{2}.
\end{equation}
The radiation density term $\rho_r $ is absent in the pre bang constraints. The cosmological parameter evolutionary equations are 
expressed in terms of the Lambert W function when possible.  

\subsection{Mathematical properties, equations and functions}\label{ss-mpef}
The equations and functions needed to describe the physical properties of the LHQ universe are developed in the following
sections.

\subsubsection{The dark energy scalar constants and current values}\label{sss-desc}
From \cite{thm23} the constants $p$ and $q$ in $\chi$ are
\begin{equation}\label{eq-qp}
p=\frac{8}{(\kappa\delta)^2} \hspace{1cm} q=-\frac{e^{\frac{c}{(\kappa\delta)^2}}}{(\kappa\delta)^2} \hspace{1cm} 
c=2(\kappa\delta)^2\ln(\kappa\theta_0)-(\kappa\theta_0)^2.
\end{equation}
and the present day scalar $\kappa\theta_0$ is
\begin{equation}\label{eq-tho}
\kappa\theta_0 =-\frac{4-\sqrt{16+12\Omega_{\theta_0}(w_0+1)(\kappa\delta)^2}}{2\sqrt{3\Omega_{\theta_0}(w_0+1)}}
\end{equation}
where $\Omega_{\theta_0}$ is the ratio of the present dark energy density to the critical density.

\begin{figure}
\includegraphics[width=8cm]{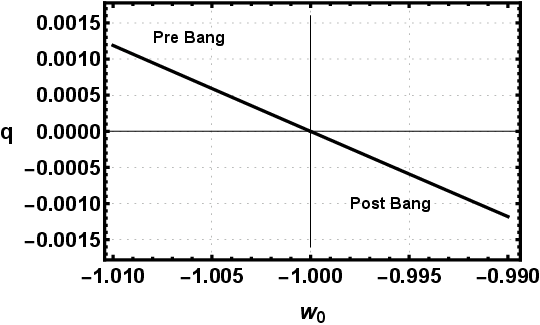}
\caption{\label{fig-q} The value of $q$ is plotted as a function of $w_0$ for pre and post bang $w_0$ values between -0.99 and 
-1.01.  The sign of $q$ transitions from negative for $w_0$ less negative than minus one to positive for $w_0$ values more negative 
than minus one.}
\end{figure}

Figure~\ref{fig-q} shows the solution for $q$ in Eqn.~\ref{eq-qp} in terms of $w_0$ using Eqns.~\ref{eq-qp} and~\ref{eq-tho}.
The post bang values of $q$ are negative which produce a negative value of $\chi$ and a real scalar.  The pre bang values are
positive making the scalar imaginary. Although both the pre bang scalar and its time derivative are imaginary all of the observable 
cosmological parameters are real. Positive values of the W function require being on the positive portion of the principal branch. 
Figure~\ref{fig-q} shows this is achieved with $w_0$ values more negative than minus one. This requirement is addressed in the 
boundary conditions for the pre bang values of $w_0$ in Sect.~\ref{s-bc}.

\subsubsection{The time derivative of the scalar}\label{sss-tds}
 The time derivative of the scalar is
\begin{equation}\label{eq-dtdt}
\dot{\theta} = \frac{d\theta}{da} \frac{da}{dt} = \frac{d\theta}{da}H a
\end{equation}
which requires an equation for $\frac{d\theta}{da}$.

The derivative of the W function $W(\chi)$ with respect to its argument $\chi$ is
\begin{equation}\label{eq-dwdx}
\frac{d W(\chi)}{d \chi} = \frac{W(\chi)}{\chi(W(\chi) + 1)}.
\end{equation}
The derivative of the scalar with respect to $\chi$ is then
\begin{equation}\label{eq-dtdx}
\frac{d \theta}{d \chi}=\frac{\delta}{2}(-W(\chi))^{-\frac{1}{2}}(-\frac{d W(\chi)}{d \chi})
\end{equation}
which  results in
\begin{equation}
\frac{d \theta}{d \chi}=\frac{\delta}{2}\frac{\sqrt{-W(\chi)}}{\chi(W(\chi)+1)} .
\end{equation}
The derivative with respect to the scale factor, using Eqn.~\ref{eq-chi}, is
\begin{equation}\label{eq-dtda}
\frac{d\theta}{da}=\frac{d \theta}{d \chi}\frac{d \chi}{da}=\frac{p\delta\sqrt{-W(\chi)}}{2a (W(\chi)+1)}
\end{equation}
where $p$ is the power of the scale factor in Eqn.~\ref{eq-chi}. The time derivative of the scalar is then
\begin{equation}\label{eq-dotth}
\dot{\theta}=\frac{p\delta\sqrt{-W(\chi)}}{2a(W(\chi)+1)}Ha=\frac{p\delta\sqrt{-W(\chi)}}{2(W(\chi)+1)}H
=\frac{4\sqrt{-W(\chi)}}{(\kappa\delta) (W(\chi)+1)}H
\end{equation}
using Eqn.~\ref{eq-qp} for the constant $p$.
It shows that both the scalar and its time derivative are zero at the singularity where the W function is zero.

\subsubsection{The HI potential constants $\delta$ and $M$}\label{sss-hipM}
The kinetic term is zero at the singularity making the dark energy EoS  $\frac{-V}{V}=-1$.  This requires a thawing post bang 
$w$ that starts at minus one and thaws to less negative than minus one at larger scale factors.  The previous study showed values 
of $\delta$ near one put the zero point in the future and produced thawing evolutions. The constant $\delta$ is therefore set at 0.95 
to satisfy the requirements. 

$M$ is a constant therefore its value can be calculated at any point on the principal branch which is taken to be a post bang scale 
factor of one. The time derivative of the scalar at $a=1$ is
\begin{equation}\label{eq-dto}
\dot{\theta}_0 = \frac{p(\kappa\delta)\sqrt{-W(\chi(1))}}{2(W(\chi(1))+1)}H_0 =\frac{4\sqrt{-W(\chi(1))}}{(\kappa\delta)
(W(\chi(1))+1)}H_0.
\end{equation}
The first Friedmann constraint at the present time is.  
\begin{equation}\label{eq-f1o}
3 H_0^2 = \frac{-8W(\chi(1))}{(\kappa\delta)^2(W(\chi(1))+1)^2}H_0^2 +(M\kappa\delta)^4(W(\chi(1))+1)^2 +\rho_{m_0} + 
\rho_{r_0}.
\end{equation}
Rearranging Eqn.~\ref{eq-f1o} gives
\begin{equation}\label{eq-f1or}
3 H_0^2 (1+\frac{8W(\chi(1))}{3(\kappa\delta)^2(W(\chi(1))+1)^2})= (M\kappa\delta)^4(W(\chi(1))+1)^2 +\rho_{m_0} + \rho_{r_0}.
\end{equation}
Solving Eqn.~\ref{eq-f1or} for $M^4$ by replacing $\dot{\theta}_0$ with Eqn.~\ref{eq-dto} and $(\kappa\theta)^2-(\kappa\delta)^2$
with Eqn.~\ref{eq-tw}  yields
\begin{equation}\label{eq-M4}
M^4= \frac{3H_0^2(1+\frac{8 W(\chi(1))}{3(\kappa\delta)^2 (W(\chi(1))+1)^2})
-\rho_{m_0} - \rho_{r_0}}{(\kappa\delta)^4(W(\chi(1))+1)^2}.
\end{equation} 
Since $\kappa\theta_0$ is a function of $w_0$ there is a different $M$ value for each of the two post bang values of $w_0$.
The calculated values of $M$ are listed in Table~\ref{tab-bc} along with the choice of $\delta$.

\subsubsection{The Hubble parameter}\label{sss-hp}
Rearranging Eqn.~\ref{eq-f1or} and  replacing $H_0$ with $H$ and $(1)$ with $(a)$ provides a solution for the post bang Hubble parameter.
\begin{equation}\label{eq-Hn}
H(a)=\sqrt{\frac{(M\kappa\delta)^4(W(\chi(a))+1)^2+\frac{\rho_{m_0}}{a^3}+\frac{\rho_{r_0}}{a^4}}{3+\frac{8W(\chi(a))}{(\kappa\delta)^2(W(\chi(a))+1)^2}}}
\end{equation}
The pre bang Hubble parameter is the same as the post bang except for the absence of the radiation density and a minus sign 
indicating contraction.

\subsubsection{The time derivative of the Hubble parameter}\label{sss-tdhp}
From \cite{thm23} the time derivative of the Hubble parameter, needed for the second Friedmann constraint, is
\begin{equation}\label{eq-hdot}
\dot{H}(a) =-\frac{1}{2}((M\kappa)^4\dot{\theta}^2+\frac{\rho_{m_0}}{a^3} +\frac{\rho_{r_0}}{a^4})
\end{equation}
where the radiation density term was added to the previous study result for the post bang epoch.  Equation~\ref{eq-dotth}
 provides the equation for $\dot{\theta}$ in terms of the W function.

\subsubsection{The kinetic term $X$}\label{sss-kX}
Equation~\ref{eq-Hn} for the Hubble parameter provides the complete equation for $\dot{\theta}$ in terms of the W function.
\begin{equation}\label{eq-tdotn}
\kappa^2\dot{\theta}=\frac{4\sqrt{-W(\chi)}}{\kappa\delta(W(\chi)+1)}\sqrt{\frac{(M\kappa\delta)^4(W(\chi(a))+1)^2+\frac{\rho_{m_0}}{a^3}+\frac{\rho_{r_0}}{a^4}}{3+\frac{8W(\chi(a))}{(\kappa\delta)^2(W(\chi(a))+1)^2}}}.
\end{equation}
Multiplying through by $\frac{1}{\kappa\delta(W(\chi(a))+1)}$ gives
\begin{equation}\label{eq-tdotnr}
\kappa^2\dot{\theta}=4\sqrt{-W(\chi)}\sqrt{\frac{(M\kappa\delta)^4(W(\chi(a))+1)^2+\frac{\rho_{m_0}}{a^3}+\frac{\rho_{r_0}}{a^4}}{3(\kappa\delta)^2(W(\chi(a))+1)^2+8W(\chi(a))}}.
\end{equation}

Using  Eqn.~\ref{eq-tdotnr} the post bang kinetic term is
\begin{equation}\label{eq-Xn}
X(a)=-8W(\chi(a))\left(\frac{(M\kappa\delta)^4(W(\chi(a))+1)^2+\frac{\rho_{m_0}}{a^3}+\frac{\rho_{r_0}}{a^4}}{3(\kappa\delta)^2(W(\chi(a))+1)^2+8W(\chi(a))}\right).
\end{equation}
The pre bang kinetic term is identical except for the absence of the radiation term. The value of $\dot{\theta}$ and the kinetic 
term are real and positive in the post bang epoch. In the pre bang era $\dot{\theta}$ is imaginary making the kinetic term negative 
and real. 

\subsubsection{The dark energy density and pressure}\label{sss-dedp}
The quintessence dark energy density and pressure evolutionary functions are given in Eqns.~\ref{eq-rhop}.   Using equation Eqn.~\ref{eq-Xn} the post bang dark energy density is
\begin{eqnarray}\label{eq-dedenn}
\rho_{de}(a)=-8W(\chi)\left(\frac{(M\kappa\delta)^4(W(\chi)+1)^2+\frac{\rho_{m_0}}{a^3}+\frac{\rho_{r_0}}{a^4}}{3(\kappa\delta)^2(W(\chi)+1)^2+8W(\chi)}\right)+(M\kappa\delta)^4(W(\chi(a))+1)^2.
\end{eqnarray}
The post bang dark energy pressure is
\begin{eqnarray}\label{eq-deprsn}
P_{de}(a)=-8W(\chi)\left(\frac{(M\kappa\delta)^4(W(\chi)+1)^2+\frac{\rho_{m_0}}{a^3}+\frac{\rho_{r_0}}{a^4}}{3(\kappa\delta)^2(W(\chi)+1)^2+8W(\chi)}\right)-(M\kappa\delta)^4(W(\chi)+1)^2
\end{eqnarray} 
The pre bang functions are identical except for the absence of the radiative term.

\subsubsection{The dark energy equation of state}\label{sss-deeos}
From the definition of the  dark energy EoS $w$ is the ratio of the dark energy pressure and density.  The dark energy EoS is
\begin{equation}\label{eq-dew}
w(a)=\frac{X(a)-(M\kappa\delta)^4(W(\chi(a))+1)^2}{X(a)+(M\kappa\delta)^4(W(\chi(a))+1)^2}
\end{equation}
where $X(a)$ are the pre and post bang kinetic terms.

\section {Boundary Conditions}\label{s-bc}
Table~\ref{tab-bc} shows the boundary conditions for the pre and post bang epochs. Most are post bang values where observational
evidence exists. Post bang $H_0=73$ (km/sec)/Mpc. The pre bang scale factor one Hubble parameter values are near, but not 
exactly 73.  They do not appear in Table~\ref{tab-bc} because they are not boundary conditions.  The two post bang $w_0$ values 
are -0.99 and -0.999, purposely set close to $\Lambda$CDM. Pre bang are -1.01 and -1.001 symmetric to post bang.  
$\Omega_{\theta_0}$ is 0.7 at $a=1$ in both epochs.  The matter and radiation ratios are 0.2999 and 0.0001 post bang 
respectively with the pre bang matter at 0.3.   

\begin{table}[h]
\caption{ \label{tab-bc}The post and pre bang boundary conditions}
\begin{tabular}{ccccc}
\hline
\multicolumn{5}{c}{The Boundary Condition and Constant Values} \\
\hline
 &\multicolumn{2}{c}{Post Bang}&\multicolumn{2}{c}{Pre Bang} \\
\hline
$H_0$ & 73 & 73 &  &  \\
\hline
$w_0$ & -0.99& -0.999  &-1.01 & -1.001\\
\hline
$\Omega_{\theta_0}$ & 0.7 & 0.7 & 0.7 & 0.7 \\
\hline
$\rho_{\theta_0}$  & 8.58555E-121 & 8.56511E-121 & 8.54043E-121 & 8.56059E-121 \\
\hline
$\Omega_{M_0}$ & 0.2999 & 0.2999 & 0.3 &  0.3 \\
\hline
$\rho_{M_0}$ & 3.6783E-121 & 3.6783E-121 & 3.67952E-121 & 3.67952E-121 \\
\hline
$\Omega_{R_0}$ & 0.0001 & 0.0001 & 0.0 & 0.0 \\
\hline
$\rho_{R_0}$& 1.22651E-124 & 1.22651E-124 & 0.0 & 0.0 \\
\hline
 $M$ & 1.01518E-30&1.01347E-30& 1.01518E-30&1.01347E-30\\
\hline
 $\delta$ & 0.95 &  0.95 &  0.95 &  0.95 \\
\hline
\end{tabular}
\end{table}
The dimensionfull units are: $H_0$, $(km/sec)/Mpc$, $\rho_{\theta_0}$, $\rho_{M_0}$ and $\rho_{M_0}$, $M_p^4$,
$M$ and $\delta$, $M_p$.  The other boundary conditions are dimensionless.

\section{The Physical Properties of the LHQ Universe}\label{s-ppc}
This section examines the evolution of cosmological parameters in the far past region and the transition zone.  The physical 
properties of the pre bang universe calculated by the mathematically deterministic functions from Sec.-\ref{ss-mpef} are presented 
in Sect.~\ref{ss-pbep} followed by the transition zone physical properties in Sect.~\ref{ss-pptz}.

\subsection{The physical properties of the pre bang epoch}\label{ss-pbep}
Although the equations and mathematical properties of the pre bang epoch are identical to the post bang the physical
properties are significantly different.  Some differences are due to the purely imaginary scalar and its derivatives. A major
difference is a contracting universe as the scalar rolls down the HI potential with a negative Hubble parameter. At the beginning
of the far past, a pre bang scale factor of 1000, the universe is cold and starless. It remains so for the entire pre bang region that 
terminates at pre bang scale factor of one. Non-dark matter is most likely elementary particles, quarks, leptons, gauge bosons and 
the Higgs boson.  The density of matter is $10^{-9}$ of the current post bang density matter density.  Dark energy dominates with 
a density more than 2000 times greater than the current  post bang density.  In the equation for the dark energy density the HI 
potential dominates the much smaller kinetic term.  This condition continues up to the present time.  The far past start scale factor 
of 1000 is deep into the region where both the W function and the scalar are significantly greater than one.  This means that the 
several $(W+1)$ terms are equal to W and the first term of the HI potential is dominant.  In the following the abscissas of the plots 
are in Log 10 of the scale factor to properly display the evolutions at all scale factors between one and 1000.  The ordinates are 
linear unless stated otherwise such as in the top panel of Fig.~\ref{fig-lwfscdp}.  

\subsubsection{Chi, the W function and the scalar}\label{sss-fpcws} 
\begin{figure}
 \centering
    \fbox{\includegraphics[width=8cm]{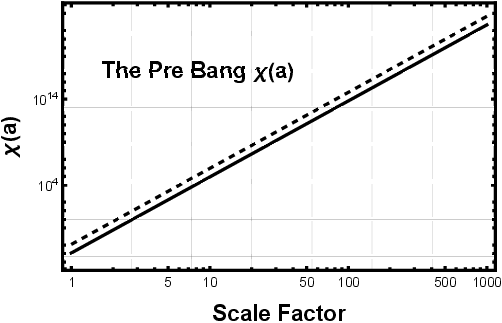}}
    \hspace{0.5cm}
    \centering
    \fbox{\includegraphics[width=8cm]{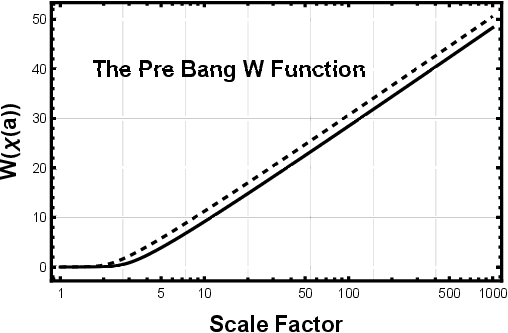}}
    \hspace{0.5cm}
    \fbox{\includegraphics[width=8cm]{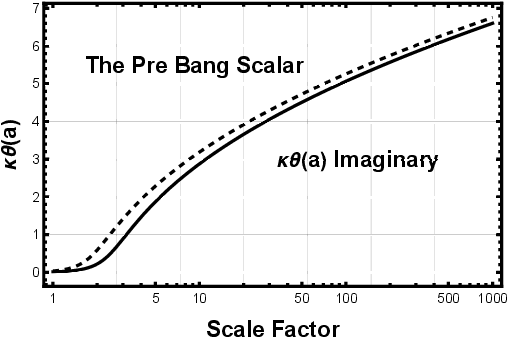}}
    \caption{The $w_0=-1.01$ cases are plotted with dashed lines and the $w_0=-1.001$ cases with solid lines. These line styles are
    used in the remainder of the pre bang plots unless stated otherwise.  The top panel shows the evolution of $\chi(a)$ in a log, log 
    plot. The other panels are log linear plots. Time evolution is from right to left.} 
    \label{fig-lwfscdp}
\end{figure}

Figure~\ref{fig-lwfscdp} shows the evolution of $\chi(a)$, the Lambert W function, and the scalar $\kappa\theta$ from a scale factor
of 1000 to 1.  The top $\chi(a)$ panel is a log log plot but the bottom two panels are the standard linear log plots.
Time evolution is from right to left with the scale factor contracting.  The Lambert W function is positive and real while 
the scalar is positive and imaginary.  The nature of the early time W function and scalar evolution persists until it nears the transition 
zone where it flattens out.  The evolution of $\chi$, $W(\chi)$ and $\kappa\theta$ are very similar for the two pre bang $w_0$
boundary conditions of -1.01 and -1.001.  All three of the functions are monotonically decreasing as the scalar rolls down the HI 
potential.  The imaginary scalar makes the square of the scalar negative instead of the post bang positive values.  The flattening 
of the evolution of the W function and the scalar near  the transition zone is an indicator that the transition zone evolution is 
atypical of the general LHQ evolution.

\subsubsection{The HI potential}\label{sss-fphi}
\begin{figure}
    \centering
    \fbox{\includegraphics[width=8cm]{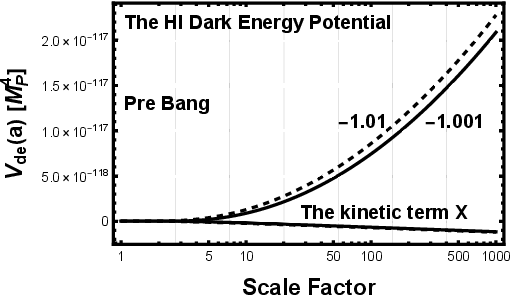}}
    \hspace{0.5cm}
    \fbox{\includegraphics[width=8cm]{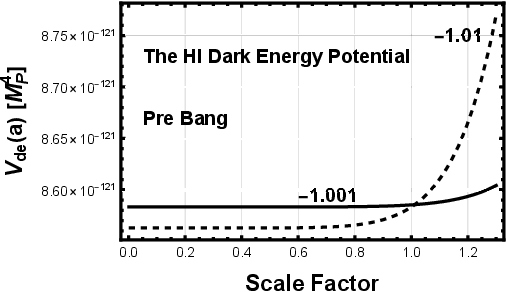}}
    \caption{The figures show the distant past evolution of the HI dark energy potential in the pre bang epoch.  The upper panel
    shows the evolution of the HI potential in the standard log linear plot. The smaller kinetic term is at the bottom of the panel. The 
    lower panel shows the detail of the cross over of the potentials near the pre bang scale factor of one.} 
    \label{fig-hipotfp}
\end{figure}

Figure~\ref{fig-hipotfp} shows the pre bang evolution of the HI potential in $(M_P)^4$.  The imaginary scalar renders all terms of
HI potential positive. The upper panel shows the evolution in the standard log linear format along with the evolution
of the kinetic term X demonstrating the dominance of the HI potential in the pre bang region.  The evolution is a smooth 
monotonic decline.  The $w_0=-1.01$ case has a slightly higher value than the $w_0=-1.001$ case.  The lower panel shows that
the $W_0=-1.01$ case crosses below the $w_0=-1.001$ case as it enters the transition zone. Comparison of the two panels indicates 
that the HI potential decreased by roughly a factor of 2300 from a pre bang scale factor of 1000 to one.  

\subsubsection{The dark energy density and pressure}\label{sss-denpres}
\begin{figure}[h!]
    \centering
    \fbox{\includegraphics[width=8cm]{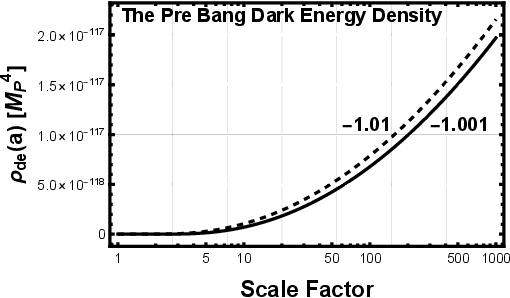}}
    \hspace{0.5cm}
    \fbox{\includegraphics[width=8cm]{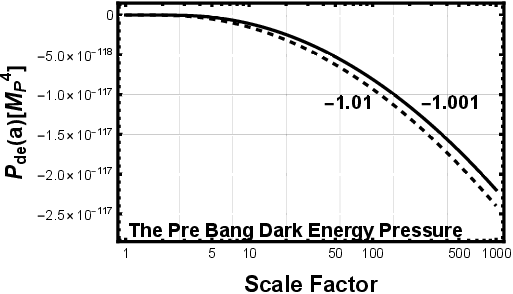}}
    \caption{The figures show the evolution of the HI dark energy density and pressure in the pre bang epoch.  
    The upper panel shows the evolution of the dark energy density and the lower panel shows the pressure.} 
    \label{fig-denpress}
\end{figure}

Figure~\ref{fig-denpress} shows the pre bang evolution of the dark energy density and pressure.  As expected, the range of evolution
is similar to that of the HI potential since both the density and pressure are potential dominated in the pre bang epoch.  The
magnitude of the pressure exceeds that of the density since both the kinetic and potential terms in the pressure are negative
while in the density the potential term is positive but the kinetic term is negative.
Both the dark energy density and pressure have almost flat evolutions as they approach the transition zone. The dark energy 
density and pressure drop by approximately a factor of $10^4$ between a scale factor of 1000 and one.

\subsubsection{The Hubble parameter}\label{sss-hub}
\begin{figure}
    \centering
    \fbox{\includegraphics[width=8cm]{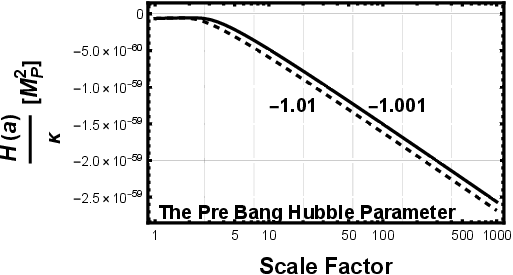}}
    \caption{The distant past evolution of the Hubble parameter in the pre bang epoch.} 
    \label{fig-Hfp}
\end{figure}

Figure~\ref{fig-Hfp} displays the evolution of the Hubble parameter which is negative in all parts of the far past.  
The Hubble parameter is dark energy dominated and becomes less negative with time following the decrease in dark energy 
density shown in the upper panel of Fig.~\ref{fig-denpress}.  The evolution is linear in the log linear plot from a scale factor of 
1000 to a scale factor of 3. The Hubble parameter evolution is essentially the negative of the W function evolution in 
Fig.~\ref{fig-lwfscdp}.  This is due to the dominance of $W(\chi)$ over one in the $(W(\chi)+1)$ terms in Eqn.~\ref{eq-Hn}. 
The pre bang Hubble parameter evolution flattens at a scale factor of 3 as it approaches the  transition zone. 

\subsubsection{The time derivative of the scalar and the kinetic term}\label{sss-dstX}
\begin{figure}
    \centering
    \fbox{\includegraphics[width=8cm]{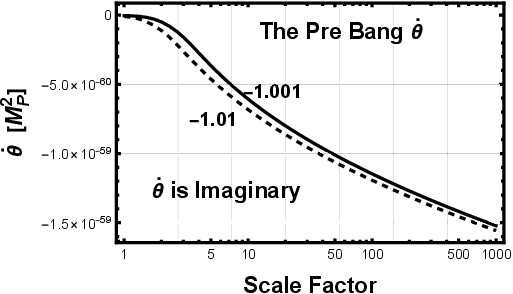}}
    \hspace{0.5cm}
    \fbox{\includegraphics[width=8cm]{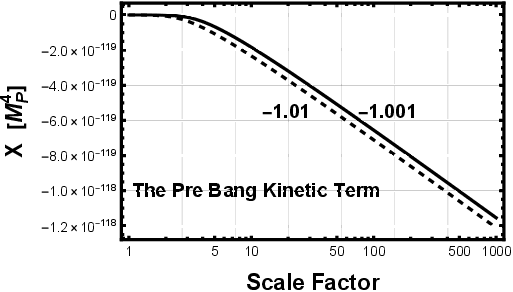}}
    \caption{The top panel shows the evolution of the time derivative of the scalar $\dot{\theta}$. The time derivative is purely
    imaginary.  The bottom panel show the kinetic term X = $\frac{\dot{\theta}^2}{2}$} 
    \label{fig-tdotX}
\end{figure}

Figure~\ref{fig-tdotX} shows the kinetic term and the time derivative of the pre bang scalar.   The pre bang time derivative of 
the scalar is negative and purely imaginary since the scalar is decreasing and purely imaginary.  It has a similar evolution as previous 
parameters with a steep decrease in magnitude and a flattening of evolution as it approaches the transition zone.  The kinetic term is 
negative but real since it is proportional to the square of an imaginary number. 

\subsubsection{The dark energy equation of state}\label{sss-fpw}
\begin{figure}
    \centering
    \fbox{\includegraphics[width=8cm]{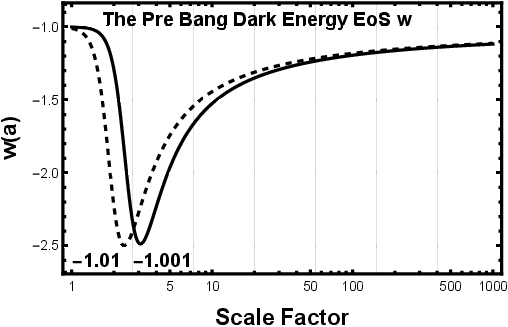}}
      \hspace{0.5cm}
     \fbox{\includegraphics[width=8cm]{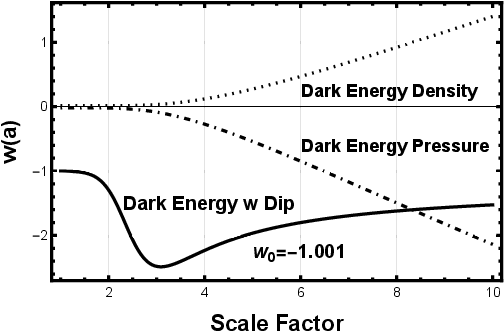}}
    \caption{ The upper panel shows the distant past evolution of the dark energy equation of state $w$. The evolution of the 
    dark energy density and pressure creating the $w$ dip is shown in the lower panel.} 
    \label{fig-wfp}
\end{figure}

The upper panel of Fig.~\ref{fig-wfp} shows the pre bang evolution of the dark energy EoS $w$.  
The most prominent feature is the negative dip at scale factors between two and three reaching $w$ values near -2.5.  The 
lower panel in Fig.~\ref{fig-wfp} shows how the dip occurs. At scale factors near the transition zone the kinetic term plays a
larger, but not dominant,  role as the HI potential declines.   The ratio of the pressure to the density reaches a maximum at a 
scale factor of three producing the dip.  At smaller scale factors the kinetic term is approaching zero making $w \approx -V/V=-1$ 
as it enters the transition zone.  The overall evolution of $w$ is from a value almost minus one at a scale factor of 1000, becoming 
more negative as it evolves to a minimum near -2.5 near a scale factor of 3 then returning toward minus one as it approaches the
transition zone.

\subsubsection{Friedmann constraints in the pre bang era}\label{ss-fcpreb}
This section examines the compliance of the derived parameters to the two Friedmann constraints, Eqns.~\ref{eq-f1} 
and~\ref{eq-f2} in the pre bang zone. The pre and post bang equations and deviations are identical except the pre bang lacks the 
radiation density in the first constraint. The equations for fractional deviations from the first, F1, and second, F2, Friedman 
constraints are
\begin{equation}\label{eq-frf1}
dev = \frac{3H^2-(\rho_{de} +\rho_m +\rho_r)}{3H^2} \hspace{0.1cm} and \hspace{0.1cm}  \frac{3(\dot{H} + H^2)+(\frac{\rho_{de} +\rho_m +\rho_r +3P}{2})}{3(\dot{H} + H^2)}.
\end{equation}

\begin{figure}
    \centering
    \fbox{\includegraphics[width=8cm]{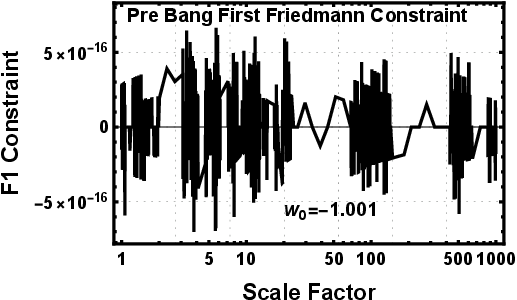}}
    \hspace{0.5cm}
    \fbox{\includegraphics[width=8cm]{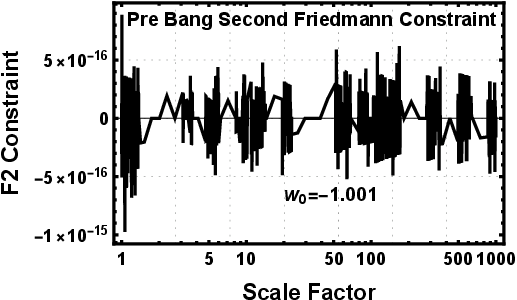}}
    \caption{The panels show the error level of the first and second Friedmann constraints for $w_0=-1.001$ case . The Friedmann constraints are satisfied to better than one part in $10^{15}$ at all points in the pre bang zone.} 
    \label{fig-f12prebang}
\end{figure}

Figure~\ref{fig-f12prebang} shows  the constraint deviations for the first and second Friedmann constraints with $w_0=-1.001$.  
Their deviations are all less than one part in $10^{15}$. The nature of the fractional errors suggest that they are the digital 
limitations of the Mathematica calculations which are on the order of a few times $10^{-16}$.
The plots of the $w_0=-1.01$ case are similar to the -1.001 case satisfying the Friedmann constraints at a high level.

\subsubsection{Pre bang chronology}\label{sss-pbc}
\begin{figure}
\includegraphics[width=8cm]{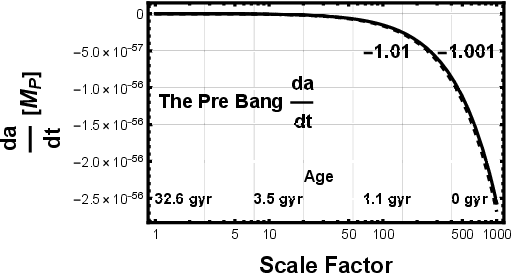}
\caption{The time derivative of the scale factor is plotted for pre bang scale factors between 1000 and 1.  The time in gigayears
for the scale factors of 1000, 100, 10, and 1 is given at the bottom of the figure.}
\label{fig-dadt}
\end{figure}

This section examines the predicted pre bang chronology.  The time derivative of the scale factor is simply $H(a)a$ for the Hubble 
parameter $H(a)$ and the scale factor $a$. The time as a function of $a$ is
\begin{equation}\label{eq-dt}
dt = \frac{da}{H(a) a}
\end{equation}
where $H(a)$ is the pre bang Hubble parameter given by the negative of Eqn.~\ref{sss-hp} without the radiation term.  The time in
gigayears from a scalar factor of 1000 to a scale factor of $a$ is calculated by numerically integrating Eqn.~\ref{eq-dt}. 
Figure~\ref{fig-dadt} shows the time derivative of the scale factor as a function of the scale factor.  The time since a scale 
factor of 1000 is shown at the bottom for scale factors of 1000, 100, 10 and 1.
The time derivative of the scale factor is of course negative in the pre bang epoch and is quite large at the pre bang scale factor
of 1000.  The contraction from a  scale factor of 1000 to 100 takes only 1.1 gigayears. The contraction to a pre bang scale
factor of 1 takes 32.6 gigayears, less than three times the age of the post bang universe.  The time to contract from a pre bang
scale factor of one to the big bang is 14.9 gigayears, slightly longer than the age of the post bang universe.  These numbers
illustrate the significant difference between the rapid contraction in the very early pre bang universe relative to the contraction
and expansion rates in the transition zone.  It further emphasizes that the transition zone is a remarkable time in the evolution
of the LHQ universe.

\subsubsection{Still a Quintessence cosmology}\label{ss-sq}
The pre bang epoch has two properties associated with a phantom cosmology: a dark energy EoS more negative than minus one 
and a negative kinetic term.  It is, however, still a quintessence cosmology rather than a phantom cosmology with well known 
instability issues \cite{cal02, car03, cli04, sch05, vik05, lud17}.  In a phantom cosmology the kinetic term is negative because it
has been changed to negative in Eqns.~\ref{eq-rhop} to give the ``wrong sign'' in the definitions of the dark energy density and
pressure \cite{bar18}. In this study the equations retain their quintessence form. The negative sign for the kinetic term is due to 
the imaginary value of $\dot{\theta}$.  Since the equations retain their quintessence form the sound speed,
$c^2_s = \frac{P,_X}{\rho,_X}$, is one in both the pre and post bang epochs.  From \cite{eri02} this indicates that the LHQ 
universe is stable in both epochs.

\subsection{The physical properties of the transition zone}\label{ss-pptz}
The transition zone contains the end of the pre bang epoch, the big bang and the entirety of the post bang observable
universe between post bang scale factors from zero to one.  It is a region of relatively little dark energy evolution but of
a highly evolving matter density.  The big bang is a natural consequence of the contraction of the pre bang universe to a
singularity.  The big bang is, therefore, no longer the origin of time and the universe. This is counter to the canonical view of 
the big bang being the origin of both time and the universe, eg. \cite{har83, haw84, har08, har13, haw14} and many others
that invoke a universe that is the result of a quantum entanglement collapsing to a given quantum state.  In LHQ the universe
is in a given state far before the big bang.  In this work our chosen scale factor of 1000 for the beginning of the evolutionary
calculations is presumed to be after the state has been established.  A quantum analysis of the establishment of the LHQ
state is beyond the scope of this investigation.  Two possibilities, however, are worth consideration.  One is that the universe
has always been in the LHQ state and does not involve a quantum entanglement.  The other possibility invokes a multiverse
option that there is a quantum entanglement that exists in the distant past before the big bang.  New universes form from the
collapse of part of these entanglements at regular intervals and we are in the one that collapsed to the LHQ state.  Neither of 
these possibilities have been examined in this work beyond pure speculation.

The big bang is, however, an essential event required to produce favorable conditions for the formation of stars and galaxies that 
is not possible in the pre bang epoch. Contraction  to a singularity and transition to a post bang universe is not a new concept 
\cite{pen65, pen69a, pen69b, gas93, kho02}.  Several studies of cyclic universes \cite{lin86, ste02a, ste02b, gao14} proposed 
universes that cycled through numerous big bangs, however, with some entropy issues due to the multiple occurrences.  The LHQ 
universe is not cyclic.  It contains only one big bang. 

\subsubsection{Expansion normalized variable examination of the transition zone}\label{sss-envet}
A successful transition from the pre bang to post bang epoch requires the universe to evolve through the big bang rather 
than bouncing back into the pre bang epoch such as described in ~\cite{ven04}.  Expansion Normalized, EN, variables 
\cite{bah18} are employed to determine whether a transition is possible.  The calculation is made near either 
side of the singularity. The quintessence kinetic $x$ and potential $y$ variables are defined in \cite{bah18} as
\begin{equation}\label{eq-env}
x =\frac{\kappa\dot{\theta}}{\sqrt{6H}} \hspace{1cm} y=\frac{\kappa\sqrt{V}}{\sqrt{3H}}.
\end{equation}
In the pre bang epoch $x$ is imaginary since $\dot{\theta}$ is imaginary but $x$ is real in the post bang epoch. The value of
$y$ is real in both epochs.

\begin{figure}
\includegraphics[width=5cm]{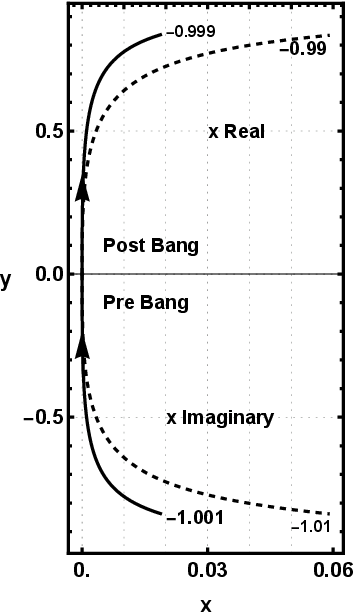}
\caption{\label{fig-env} The evolution of the EN variables $x$ and $y$ at the pre and post bang boundary.  The
arrows indicated the direction of the evolution.  The numbers at the ends of the track indicate the appropriate
$w_0$ values for the track.}
\end{figure}

Figure~\ref{fig-env} shows the evolution of the EN variables in the transition zone. 
It indicates that a transition through the singularity from the pre to post bang epoch is possible.  The figure also shows the 
dominance of the $y$ potential variable over the kinetic $x$ variable.  Even though the dark energy potential $V$ does not go to 
zero the $y$ value is zero at the singularity due to the infinite Hubble parameter there. The kinetic evolution is near zero on both 
sides of the singularity for a significant portion of the plot, particularly for the $w_0$ value closest to minus one.  

\subsubsection{$\chi(a)$, the Lambert W function and the scalar}\label{sss-tzwt}
The flatness of the kinetic term evolution in the EN plot is an indicator of a region of suppressed evolution of dark energy on 
either side of the singularity.  It is due to the slow evolution of $\chi(a)$ for scale factors less than one.  Equation~\ref{eq-qp} for 
$\chi(q a^p)$ shows that the power of the scale factor is larger than 8 which dampens the evolution for scale factors less than one. 
 
\begin{figure}
    \centering
    \fbox{\includegraphics[width=8cm]{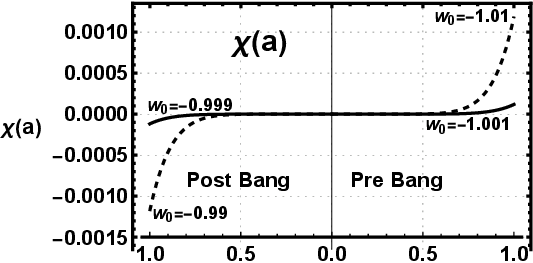}}
    \hspace{0.5cm}
    \fbox{\includegraphics[width=8cm]{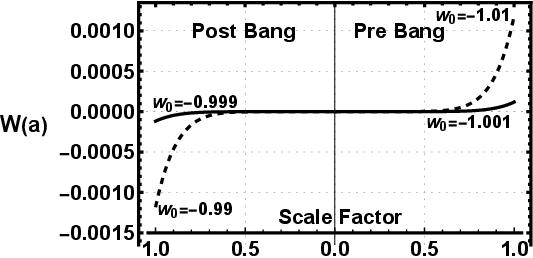}}
    \hspace{0.5cm}
    \fbox{\includegraphics[width=8cm]{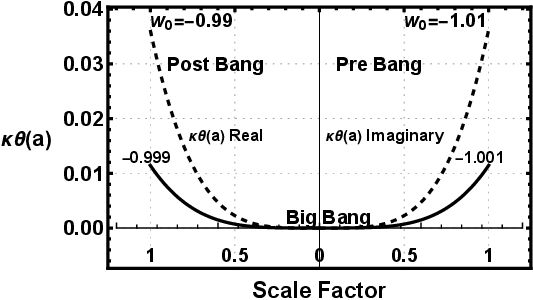}}
    \caption{The upper panel shows the evolution of $\chi(a)$ through the transition zone. As was the case in the pre bang plots
    the $w_0=-0.999$ case is the solid line and the $w_0=-0.99$ case is the dashed line.  The evolution is from 
    the pre to post bang epoch, right to left. The middle panel shows the evolution of the W function and the bottom panel
    the evolution of the scalar.} 
    \label{fig-tzcws}
\end{figure}

The dampening of the evolution of $\chi(a)$ is apparent in the upper panel of Fig.~\ref{fig-tzcws}. The middle panel of 
Fig.~\ref{fig-tzcws} shows the Lambert W function evolution and the bottom panel the evolution of the scalar $\kappa\theta$.  
The scalar has a flat evolution between scale factors of 0.5 on either side of the singularity.  The $w_0=-0.99$ case has significantly 
more evolution than the $w_0=-0.999$ case  showing the effect of $w_0$ on evolution.  All three of the parameters in 
Fig.~\ref{fig-tzcws} are zero at the singularity.  The change of the scalar from imaginary to real is an important aspect of the
transition.  The scalar is much smaller than one throughout the transition zone, particularly for the $w_0$ case closest to
minus one.  This is a significant factor in the evolution of the HI potential discussed in the next section. 

\subsubsection{The HI dark energy potential}\label{ss-tzhip}
\begin{figure}
\includegraphics[width=8cm]{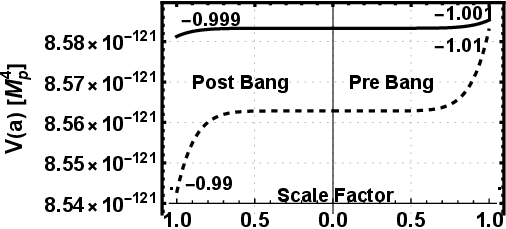}
\caption{\label{fig-tzhi} The figure shows the evolution of the HI potential, Eqn.~\ref{eq-HIpot}, in the transition zone
for the two different $w_0$ cases. Compared to their absolute values there is relatively insignificant evolution of the HI potential in 
the transition zone.}
\end{figure}

At the singularity the HI dark energy potential is equal to the constant term $(M\kappa\delta)^4$  since the scalar $\kappa\theta=0$.  
Figure~\ref{fig-tzhi} shows that the evolution of the HI potential is essentially flat in the transition zone.
The $w_0=-0.999$ case deviation is only $0.1\%$ at a scale factor of one.  The slow evolution occurs because the scalar remains
much less than one in the transition zone as shown in the lower panel of Fig.~\ref{fig-tzcws}. This makes the first two terms of
the HI potential in Eqn.~\ref{eq-HIpot} much smaller than the constant  third term. Figure~\ref{fig-tzhi} shows that the HI potential
continues to decrease as it exits the transition zone.  The decrease is due to the second term of HI potential changing from positive
in the pre bang epoch to negative in the post bang era as the scalar changes from imaginary to real.  The flatness of the dark energy
potential evolution in the transition zone is a combination of the scalar being almost zero near the singularity due to the W function, 
the presence of the constant term in the HI potential, and the dampening described in Sect.~\ref{sss-tzwt}.  As discussed in 
Sect.~\ref{sss-tzdedp} this leads to an evolution of cosmological parameters in the transition zone very close to $\Lambda$CDM with 
no fine tuning.

\subsubsection{The Hubble parameter}\label{sss-tzH}
\begin{figure}
    \centering
    \fbox{\includegraphics[width=8cm]{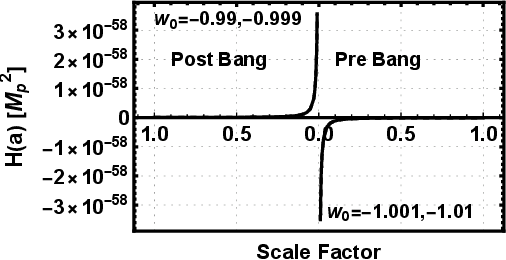}}
    \hspace{0.5cm}
    \fbox{\includegraphics[width=8cm]{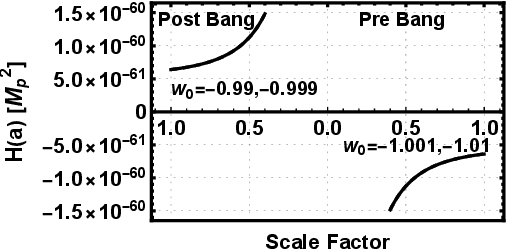}}
    \caption{The upper panel shows the transition zone evolution of the Hubble parameter from scale factors between 0.01 and 1.0.
    in units of $M_P^2$ on either side of the singularity.  The lower panel is for scale factors between 0.4 and 1.0.  The tracks for 
    the two $w_0$ cases are the same to the width of the plot line but not identical.} 
    \label{fig-tranH}
\end{figure}

The negative pre bang and positive post bang Hubble parameters create the expected discontinuity marking the transition from a 
contracting to an expanding universe at the singularity.  Figure~\ref{fig-tranH} shows the transition zone Hubble parameter evolution 
near the singularity.  The upper panel tracks terminate at a scale factors of 0.01 to show the discontinuity.  The lower panel tracks 
terminate at scale factors of 0.4 to better show the evolution at larger scale factors
The tracks for the two different $w_0$ values overlap at the resolution of the plot. This demonstrates the insensitivity of the 
Hubble parameter to $w_0$, consistent with previous studies \cite{thm18, thm19}.

\begin{figure}
\includegraphics[width=8cm]{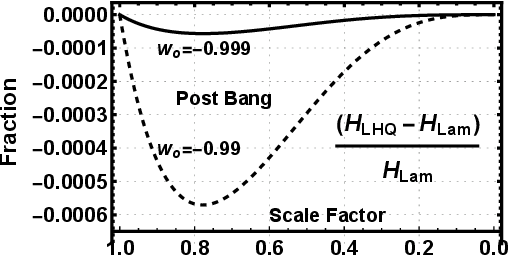}
\caption{\label{fig-fracH} The figure shows the fractional deviation of the two LHQ Hubble parameter evolutions from the
$\Lambda$CDM evolution for the post bang era. It is plotted with time evolution from right to left to be consistent with the 
previous plots.}
\end{figure}

Figure~\ref{fig-fracH} shows the fractional deviation of the post bang Hubble parameter from $\Lambda$CDM , defined as 
$\frac{(H_{LHQ} - H_{\Lambda})}{H_{\Lambda}}$, for the two $w_0$ values.
The maximum fractional deviation is 0.006 for the $w_0= -0.99$ case and only 0.00005 for the $w_0=-0.999$ fiducial case.  The
Hubble parameter is not a sensitive discriminator between static and dynamical cosmologies for $w_0$ values close to minus one.
This makes it hard to verify $\Lambda$CDM or falsify LHQ based solely on observations of the Hubble parameter.

\subsubsection{The time derivative of the scalar and the kinetic term.}\label{sss-tztdtx}
\begin{figure}
    \centering
    \fbox{\includegraphics[width=8cm]{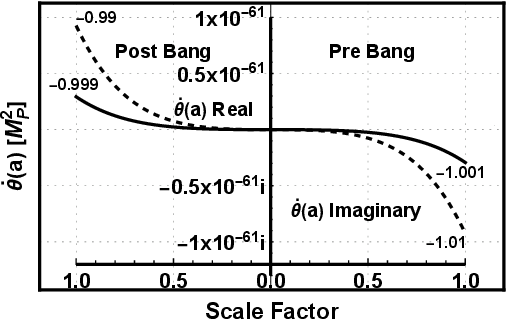}}
    \hspace{0.5cm}
    \fbox{\includegraphics[width=8cm]{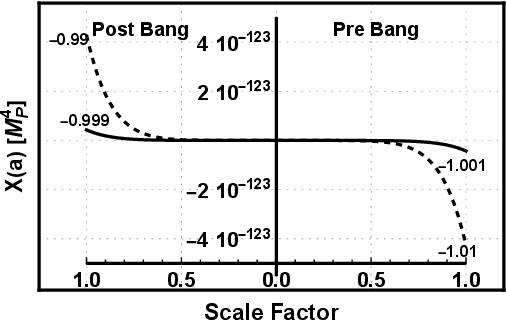}}
    \caption{The top panel shows the transition zone evolution of the time derivative of the scalar $\dot{\theta}$. 
It is negative and imaginary in the pre bang epoch and transitions to positive and real in the
post bang epoch.  The bottom panel shows the evolution of the kinetic term $X$ which is negative in the pre bang epoch and
positive in the post bang epoch.}
    \label{fig-xtdottran}
\end{figure}

The upper panel of Fig.~\ref{fig-xtdottran} shows the evolution of the time derivative of the scalar $\dot{\theta}$ in units of 
$M_P^2$ and the lower panel shows the evolution of the kinetic term $X=\frac{\dot{\theta}^2}{2}$ in units of  $M_P^4$.  
Like the scalar, $\dot{\theta}$ is zero at the transition and stays near zero for scale factors of $\approx0.5$ on either side of the transition, particularly for the $w_0=-0.999$ case. $\dot{\theta}$ is negative and imaginary in the pre bang epoch.  It
becomes real and positive at the big bang transition point.

The kinetic term $X$ evolution is quite flat in the transition zone.  This is consistent with the dominance of the potential in the EN 
variables in Fig.~\ref{fig-env}. Its value is real and negative in the pre bang epoch since it is the square of an imaginary number.

\subsubsection{The dark energy density and pressure}\label{sss-tzdedp}
\begin{figure}
    \centering
    \fbox{\includegraphics[width=8cm]{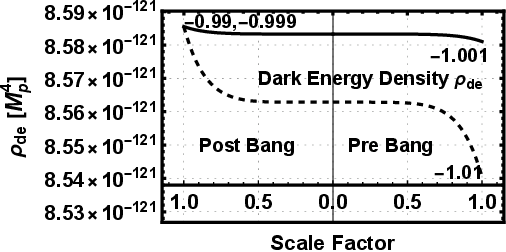}}
    \hspace{0.5cm}
    \fbox{\includegraphics[width=8cm]{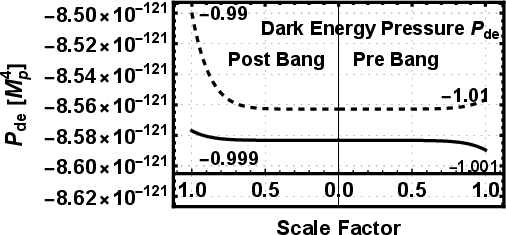}}
    \caption{The top panel shows the evolution of the dark energy density in the transition zone. The bottom panel shows the 
    evolution of the dark energy pressure. The post bang dark energy scale factor one densities are identical because it is a
    boundary condition.}
    \label{fig-dedptran}
\end{figure}

Figure~\ref{fig-dedptran} shows the evolution of the dark energy density and pressure. 
The dark energy densities for both $w_0$ values are identical at a post bang scale factor of one since the density is a boundary 
condition.  Both the dark energy density and dark energy pressure have smooth transitions from the pre bang to post bang era.
Their evolutions are almost flat through the transition zone but are not zero due to the constant in the HI potential.  The slight
increase in the dark energy density in the post bang era is primarily due to an increase in the kinetic term $X$ as shown in 
fig.~\ref{fig-xtdottran}.

The almost constant dark energy density in the transition zone, particularly for the $w_0=-0.999$ case, acts like a cosmological
constant.  It means that LHQ has essentially the same success in matching the observations as $\Lambda$CDM in the post bang
observable universe.  A large part of this success is due to the $w_0$ values very close to minus one.  Much larger deviations from 
minus one would not be as successful as shown by the higher deviation of the $w_0=-0.99$ case.  On the other hand if observational 
constraints become much more rigorous favoring, $\Lambda$CDM,  the LHQ cosmology can match them by decreasing the deviation 
of $w_0$ from minus one.  Even though the match to the observations may be the same the scalar field source of dark energy in 
LHQ is profoundly different than the cosmological constant source in $\Lambda$CDM.  The very different properties of the two 
universes outside of the transition zone are also profound.  

\subsubsection{The dark energy equation of state}\label{sss-w}
\begin{figure}
\includegraphics[width=8cm]{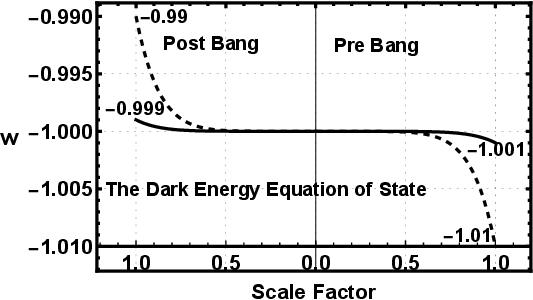}
\caption{\label{fig-deos} The evolution of the dark energy equation of state $w$ in the transition zone.}
\end{figure}

Figure~\ref{fig-deos} shows the evolution of $w$ in the transition zone.
It shows that $w$ is or is very close to minus one between the pre and post bang scale
factors of 0.5 on either side of the singularity.  This corresponds to the flat evolution of the dark energy density and pressure
between the same scale factors in Fig.~\ref{fig-dedptran}. In the post bang era $w$ demonstrates a thawing evolution.   The 
$w_0=-0.999$ case $w$ does not deviate from minus one until very recent times and even then it would be difficult to reliably 
distinguish it from minus one.  Confirmation of such a deviation of $w$ from minus one would negate $\Lambda$CDM and be
significant evidence for a dynamical cosmology such as LHQ.  The LHQ predictions for the evolutions of the observables such
as the Hubble Parameter and the dark energy EoS meet all of the current observational constraints.

\subsubsection{Friedmann constraints in the transition zone}\label{sss-f1f2check}

\begin{figure}
    \centering
    \fbox{\includegraphics[width=8cm]{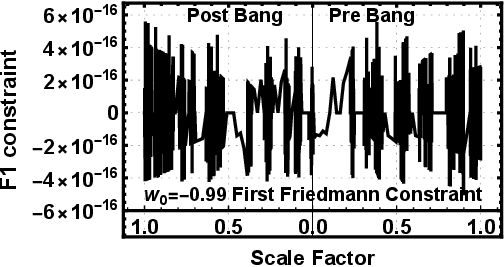}}
    \hspace{0.5cm}
    \fbox{\includegraphics[width=8cm]{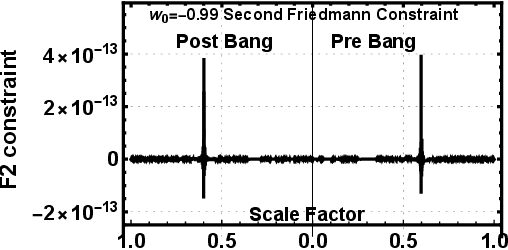}}
    \caption{The panels show the error level of the first Friedmann constraint for $w_0=-0.99$ case in the top panel
    and the second Friedmann constraint in the bottom panel.} 
    \label{fig-f12tran}
\end{figure}

Figure~\ref{fig-f12tran} shows the fractional deviation for the $w_0=-0.99$ case of the first Friedmann constraint in the 
upper panel and the second Friedmann constraint in the lower panel.
All of the fractional deviations are less than $10^{-15}$ for the first Friedmann constraint indicating that it is satisfied. 
The spikes in the second Friedmann constraint plots are caused by the denominator in Eqn.~\ref{eq-frf1} becoming zero when 
$\dot{H}$ and $H^2$ are equal but with opposite signs.  The plots are discrete points therefore the height of  the spikes depends 
on how close the discrete plot point is to the zero point.  The plot away from the spikes is similar to the first Friedmann constraint 
plot. Even with the spikes the second Friedmann constraint is also well satisfied.  The plots for the $w_0=-0.999$ case are similar.

\section{Conclusions}\label{s-con}
LHQ is a self consistent, analytic set of deterministic cosmological parameter evolutions. The evolutions are mathematically valid at 
all points on the Lambert W function principal branch.   The principal branch spans from the infinite past to a terminal point in the 
future.  LHQ has two distinct epochs, the pre big bang epoch and the post big bang epoch.  The post bang epoch, from the big 
bang to the present time is a canonical quintessence universe with a quartic polynomial dark energy potential.  As shown in
fig.~\ref{fig-W} it is after the big bang at $\chi$ and $W(\chi)$ of zero where both $\chi$ and $W(\chi)$ are negative resulting 
in a real and positive scalar.  The bottom panel of fig.~\ref{fig-W} shows that the entire canonical universe only occupies a
minute region compared to the total span of the W function principal branch.

The pre bang epoch spans from the infinite past to the big bang singularity, a vast space on the principal branch relative to the
post bang space.  It raises the possibility of the universe existing before the big bang.  The big bang is, in fact, a natural 
consequence of the contraction of the pre bang universe to the big bang singularity.  In order to explore a possible pre bang
universe this work assumes that the mathematically valid evolutions in the pre bang epoch are physically valid as well. Under
this assumption the calculated cosmological parameter evolutions reveal a noncanonical universe that is greatly, perhaps
infinitely, older and at early times larger, than the canonical post bang  universe.  The evolution of the total LHQ universe starts
with a very low matter density, high dark energy density and vastly larger universe that is contracting rather than expanding.
Its basic equations are the same quintessence equations of the post bang universe but it is cold, starless, and has a scalar
whose value is imaginary rather than real.  It is rolling down the HI potential reaching a matter singularity at a scale factor of zero.
The values of $\chi$, $W(\chi)$ and the scalar $\kappa\theta$ are also zero but the HI potential is not.  Since the scalar is zero the
HI potential is equal to its constant term $M^4(\kappa\delta)^4$ producing a constant dark energy density similar to
$\Lambda$CDM at the instant of the big bang.  The HI potential is proportional to $(W(\chi(a))+1)^2$, eqn.~\ref{eq-HIWpot},
therefore it takes a long time for $W(\chi(a)))$ to increase enough to significantly alter the HI potential. This makes the LHQ 
universe similar to $\Lambda$CDM in most of the transition zone.

The $\Lambda$CDM like behavior in the transition zone is not achieved by fine tuning but occurs due to the selection of two
important arguments, the HI potential constant term $\kappa\delta$ and the dark energy EoS boundary values $w_0$.  The
$\Lambda$CDM dark energy EoS is a constant $w=-1$.  In \cite{thm23} is was found that $\kappa\delta$ values near one
produced thawing dark energy EoS evolutions with $w=-1$ at a scale factor of zero, hence the choice of $\kappa\delta=-0.95$.
A goal of the study was to provide post bang analytic evolutions close to $\Lambda$CDM for use in likelihood comparisons 
between $\Lambda$CDM and LHQ therefore the $w_0$ values are set close to minus one, -0.99, -0.999, -1.001 and -1.01.
This limits the evolution of $w$ in the transition zone to be between minus one and the boundary condition $w_0$ values.

The canonical post bang LHQ universe has significant flexibility as a dynamical cosmology template for comparison to 
observations.  The investigation in \cite{thm23} determined that dark energy EoS evolution can be thawing, as in 
the case here, thawing but transition to freezing, or freezing only, by changing the value of the constant $\kappa\delta$ in the
HI potential to be near one, two or three respectively. This provides the means to determine the relative
likelihoods over a wide range of EoS evolutions.  Varying the present day boundary conditions on $w_0$ increases that range.
The same is true for all of the boundary conditions listed in Table~\ref{tab-bc}.  This is one of the significant advantages of
having analytic equations as a function of the scale factor for the evolutions of cosmological parameters as opposed to numerical
solutions that do not provide the same insights as analytic equations.

Although not a subject of the present investigation it is worthwhile to mention one aspect of the future evolution of the LHQ
universe.  The termination of the Lambert W function principal branch at $\chi(a)=-\frac{1}{e}$ is the minimum and zero point 
of the HI potential.  Any evolution onto the minus branch requires a reduction of the scale factor and an increase in the magnitude
of the HI potential.  As such the end of the principal branch may be a benign stable equilibrium point for the LHQ universe.  This 
would provide a future with no big rip or crunch.  This possibility will be the subject of a subsequent investigation.

The LHQ universe is an important addition to a list of cosmologies for comparison with current and the expected future
cosmological observational data.  It provides a noncanonical possibility of a universe that exists before the big bang which is
far older and once much larger than the canonical $\Lambda$CDM or post bang LHQ universes.  Exciting new observations
deserve investigations that are outside of the box if we are to discover new aspects of cosmological physics not present in our
current canonical universe models.

\acknowledgments{The author acknowledges the very useful and informative discussions with Sergey Cherkis on the mathematical properties of the Lambert W function.}

\funding{This research received no external funding} 

\dataavailability{No new data was produced in this study.}

\conflictsofinterest{The author declares no conflicts of interest.}

\begin{adjustwidth}{-\extralength}{0cm}

\reftitle{References}

\end{adjustwidth}

\begin{thebibliography}{}

\bibitem[Ratra and Peebles(1988)]{rat88} Ratra, B. and Peebles, P. 1988, Physical Review D, 37, 12
\bibitem[Peebles and Ratra(1988)]{pee88} Peebles, P. and Ratra, B. 1988, The Astrophysical Journal, 325, L17
\bibitem[Thompson(2023)]{thm23} Thompson, R. I. 2023, Universe, 9(4), 172
\bibitem[Olver et al.(2010)]{olv10} Olver, F.W.J.,  Lozier, D. W., Boisvert,
	R.F., and Clark, C.W.,  2010, in NIST Handbook of Mathematical Functions, Chap. 4, p. 111, 1st ed, (Cambridge University 
	Press, New York)
\bibitem[Desi Collaboration(2024)]{des24} Desi Collaboration 2024 arXiv:2024.03002 V1
\bibitem[Mukherjee and Sen(2024)]{muk24} Mukherjee, P. and Sen, A. A. 2024 arXiv.2405.19178v1
\bibitem[Roy(2024)]{roy24} Roy, N. 2024 arXiv:2406.00634v1
\bibitem[Jia, Hu and Wang(2024)]{jia24} Jia, X.D., Hu, J.P., and Wang, F.Y. 2024 arXiv:2406.02019v1
\bibitem[Orchard and Cardenas(2024)]{orc24} Orchard, L. and Cardenas, V. H. 2024 arXiv:2407.05579v1
\bibitem[Pourojaghi, Malekjani and Davari]{pou24} Pourojaghi, S., Mlekjani, M. and Davari Z. arXiv:2407.09767v1
\bibitem[Giare et al.(2024)]{gia24} Giare, W., Najafi, M., Pan, S., Di Valentino, E., and Firouzjaee, J. T. arXiv:2407.16689v1
\bibitem[Dinda and Maartens(2024)]{din24} Dinda, B. R., and Maartens, R. arXiv:2407.17252v1
\bibitem[Caldwell(2002)]{cal02} Caldwll, R. R. 2002, Phys. Lett. B, 545, 23
\bibitem[Carroll, Hoffman and Troden(2003)]{car03} Carroll, S. M., Hoffman, M. and Trodden, M. 2003, Phys. Rev. D, 68, 023509
\bibitem[Cline, Jeon and Moore(2004)]{cli04} Cline, J. M., Jeon, S. Y. and Moore, G. D. 2004, Phys. Rev. D, 70, 043543
\bibitem[Scherrer(2005)]{sch05} Scherrer, R. J. 2005, Phys. Rev. D, 71, 063519
\bibitem[Vikman(2005)]{vik05} Vikman, A. 2005, Phys. Rev. D, 71, 023525
\bibitem[Ludwick(2017)]{lud17} Ludwick, K. J. 2017,  Modern Physics Letters A, 32, 1730025
\bibitem[Barenboim, G., et al.(2018)]{bar18} Barenboim, G., Kinney, W. H. and Morse, M. J. P. 2018, Phys. Rev. D. 98, 083531
\bibitem[Erickson, Caldwell, Steinhardt, Armendariz-Picon and Mukhanov(2002)]{eri02} Erickson, J. K.,
	Caldwell, R., Steinhardt, P. J., Armendariz-Picon, C. and Mukhanov, V. 2002, Physical Review Letters, 88, 121301
\bibitem[Hartle and Hawking(1983)]{har83} Hartle, J. B. and Hawking, S. W. 1983, Phys. Rev. D, 28, 2960
\bibitem[Hawking(1984)]{haw84} Hawking, S. W. 1984, Nucl. Phys. B, 239, 257
\bibitem[Hartle, Hawking and Hertog(2008)]{har08} Hartle, J. B., Hawking, S. W. and Hertog, T. 2008, Phys. Rev. D, 77, 123537
\bibitem[Hartle and Hertog(2013)]{har13} Hartle and Hertog, T. 2013, Phys. Rev. D, 88, 123516
\bibitem[Hawking(2014)]{haw14} Hawking, S. W. 2014, European Physics H, 39, 413
\bibitem[Penrose(1965)]{pen65} Penrose, R., 1965, Phys. Rev. Lett., 14, 57
\bibitem[Penrose(1969a)]{pen69a} Penrose, R., 1969a, Contemporary Physics, 1, 545.
\bibitem[Penrose(1969b)]{pen69b} Penrose, R., 1969b, Rivista del Nuovo Dimento Numero Speziale, 1, 252
\bibitem[Gasperini and Veneziano(1993)]{gas93} Gasperini, M. and Veneziano, G., 1993, Astropart Phys. 1, 317
\bibitem[Khoury et al.(2002)]{kho02} Khoury, J., Ovrut, B. A., Seiberg, N., Steinhardt, P. J. and Turok, N.,
	2002, Phys. Rev. D, 65, 086007-1,8
\bibitem[Linde(1986)]{lin86} Linde, A. D., 1986, Physics Letters B, 175, 395
\bibitem[Steinhardt and Turok(2002a)]{ste02a} Steinhardt, P. J. and Turok, N. 2002, Physical Review D, 65, 126003
\bibitem[Steinhardt and Turok(2002b)]{ste02b} Steinhardt, P. J. and Turok, N. 2002, Science, 296, 1436
\bibitem[Gao, Lu and Shen(2014)]{gao14} Gao, C., Lu, Y. and Shen, Y.-G. 2014, General Relativity and Gravity, 46, 1791
\bibitem[Veneziano(2004)]{ven04} Veneziano, G. 2004, Journal of Cosmology and Astroparitcle Physics, 03, 004
\bibitem[Bahamonde et al.(2018)]{bah18} Bahamonde, S., Bohmer, C., Carloni, S.,Copeland, E.,
	Fang, W., and Tamanini, N. 2018, Physics Reports, 775-777, 1-122 
\bibitem[Thompson(2018)]{thm18} Thompson, R.I., 2018,  MNRAS, 477, 4104.
\bibitem[Thompson(2019)]{thm19} Thompson, R.I., 2019,  MNRAS, 482,  5448.
\end{thebibliography}
\end{document}